\pdfoutput=1
\documentclass[acmsmall,10pt]{acmart}
\usepackage[utf8]{inputenc}
\usepackage[subpreambles=false]{standalone}
\usepackage{xcolor}
\usepackage{mathpartir}
\usepackage{bcprules}
\usepackage{tcolorbox}
\usepackage{import}
\usepackage{multicol}
\usepackage{listings}
\usepackage{listings-rust}
\usepackage{listings-scheme}

\usepackage{xspace}

\providecommand{\sub}{}
\providecommand{\comment}{}

\providecommand{\Fm}{}
\providecommand{\judgement}[2]{}
\providecommand{\box}{}
\providecommand{\ruledef}[1]{}
\providecommand{\ruledefN}[2]{}
\providecommand{\ruledefR}[2]{}
\providecommand{\ruleref}[1]{}
\providecommand{\rulerefN}[2]{}
\providecommand{\gap}{}

\providecommand{\readonly}{}
\providecommand{\comment}{}

\providecommand{\seal}{}
\providecommand{\highlight}[1]{}
\providecommand{\reduces}{}
\providecommand{\Fsub}{}
\providecommand{\ruleref}[1]{}
\providecommand{\Lm}{}

\renewcommand{\comment}[1]{}

\renewcommand{\sub}{~\texttt{<:}~}
\renewcommand{\Fm}{System $\texttt{F}_{\texttt{<:M}}$\xspace}
\renewcommand{\Fsub}{System $\texttt{F}_{\texttt{<:}}$\xspace}
\renewcommand{\Lm}{System $\lambda_{\texttt{M}}$\xspace}

\renewcommand{\judgement}[2]{{\bf\textsf{#1}} \hfill #2}
\newcounter{RuleRef}
\renewcommand{\ruledef}[1]{%
  \refstepcounter{RuleRef}\textsc{#1}\label{rule:#1}}
\renewcommand{\ruledefN}[2]{%
  \refstepcounter{RuleRef}\textsc{#2}\label{rule:#1}}
\renewcommand{\ruledefR}[2]{%
  \refstepcounter{RuleRef}#2\label{rule:#1}}
\renewcommand{\ruleref}[1]{\textsc{(\hyperref[rule:#1]{#1})}}
\renewcommand{\rulerefN}[2]{\textsc{(\hyperref[rule:#1]{#2})}}
\renewcommand{\gap}{\quad\quad}

\renewcommand{\readonly}{\texttt{readonly}~}
\definecolor{light-gray}{gray}{0.95} 
\renewcommand{\highlight}[1]{\colorbox{light-gray}{$\displaystyle #1$}}

\renewcommand{\seal}{\texttt{seal}~}
\renewcommand{\reduces}{\;\longrightarrow\;}
\renewcommand{\ruleref}[1]{\textsc{(\hyperref[rule:#1]{#1})}}

\acmJournal{PACMPL}
\acmVolume{X}
\acmNumber{OOPSLA2}
\acmArticle{1}
\acmYear{2023}
\acmMonth{10}
\acmDOI{} 
\setcopyright{none}
\startPage{1}

\begin{document}
\title{Simple Reference Immutability for \Fsub}
\subtitle{Keeps objects fresh for up to 5X longer!}

\author{Edward Lee}
\affiliation{
   \department{Computer Science}              
   \institution{University of Waterloo}            
   \streetaddress{200 University Ave W.}
   \city{Waterloo}                                                     
   \state{ON}
   \postcode{N2L 3G1}
   \country{Canada}                    
 }
\author{Ondřej Lhoták}
\affiliation{
   \department{Computer Science}              
   \institution{University of Waterloo}            
   \streetaddress{200 University Ave W.}
   \city{Waterloo}                                                     
   \state{ON}
   \postcode{N2L 3G1}
   \country{Canada}                    
 }
\date{October 2023}

\begin{abstract}
    Reference immutability is a type based technique for taming mutation that has long been
    studied in the context of object-oriented languages, like Java.  Recently, though,
    languages like Scala have blurred the lines between functional programming languages
    and object oriented programming languages.  We explore how reference immutability
    interacts with features commonly found in these hybrid languages, in particular
    with higher-order functions -- polymorphism -- and subtyping.  We construct a calculus \Fm
    which encodes a reference immutability system as a simple extension of \Fsub and prove
    that it satisfies the standard soundness and immutability safety properties.
\end{abstract}



\maketitle

\section{Introduction}
Code written in a pure, functional language is {\it referentially transparent} -- it has no side
effects and hence can be run multiple times to produce the same result.  Reasoning about
referentially transparent code is easier for both humans and computers. However, purely
functional code can be hard to write and inefficient, so many functional languages contain
impure language features.

One important side effect that is difficult to reason about is {\it mutation} of state.
Mutation arises naturally, but can cause bugs which can be hard to untangle;
for example, two modules which at first glance are completely unrelated may interact
through some shared mutable variable.  Taming -- or controlling -- where
and how mutation can occur can reduce these issues.

One method of taming mutation is {\it reference immutability} \cite{10.1145/1103845.1094828,10.1145/2384616.2384680}.
In this setting, the type of each reference to a value can be either mutable or immutable.
An immutable reference cannot be used to mutate the value or any other
values transitively reached from it.

Mutable and immutable references can coexist for the same value, so an immutable reference
does not guarantee that the value will not change through some other,
mutable reference. This is in contrast to the stronger guarantee of {\it object immutability},
which applies to values, and ensures that a particular value does not change through
any of the references to it.

Reference immutability has long been studied in existing object-oriented programming
languages such as Java \cite{10.1145/1103845.1094828,10.1145/2384616.2384680,10.1145/1287624.1287637} 
and C\# \cite{10.1145/2384616.2384619}.  However, reference immutability is largely unexplored
in the context of functional languages with impure fragments -- languages like Scala or OCaml, for example.
Many programs in Scala are mostly immutable \cite{Haller_2017}.
A system that formally enforces specified patterns of immutability would help programmers and compilers
better reason about immutability in such programs.

One feature that is important in all languages but especially essential in functional programs is polymorphism.
The interaction of polymorphism and reference immutability raises interesting questions.
Should type variables abstract over annotated types including their immutability annotations (such as \texttt{@readonly String}), or only
over the base types without immutability annotations (such as \texttt{String})? Should uses of type variables
admit an immutability annotation like other types do? For example, should \texttt{@readonly X} be allowed, where \texttt{X} is a type variable rather than a concrete type?
If yes, then how should one interpret an annotated variable itself instantiated with an annotated type?
For example, what should the type \texttt{@readonly X} mean if the variable \texttt{X} is instantiated with \texttt{@mutable String}?

Our contribution to this area is a {\it simple} and {\it sound} treatment of reference immutability in \Fsub~\cite{DBLP:conf/tacs/CardelliMMS91}.
Specifically, we formulate a simple extension \Fm of \Fsub with the following properties:
\begin{itemize}
    \item {\bf Immutability safety:}  When dealing with reference immutability, one important property
          to show is {\it immutability safety}: showing that when a reference is given a read-only type, then the underlying value is not modified through that reference.  
          In \Fm we introduce a dynamic form of immutability, a term-level $\seal$ construct, which makes precise the
          runtime guarantees that we expect from a reference that is statically designated as immutable by the type system.
          We do this by formalizing \Lm, an untyped calculus with references and seals.
          Dynamic seals are transitive in that they seal any new references that are read from a field of an object through
          a sealed reference.
    \item {\bf \Fsub-style polymorphism:} \Fm preserves the same bounded-quantification
        structure of \Fsub. At the same time, it allows type variables to be further modified by immutability
        modifiers. 
    \item {\bf Immutable types are types:} To allow for \Fsub-style polymorphism, we need to treat immutable types
        as types themselves.  To do so, instead of type qualifiers, we introduce a type \emph{operator} $\readonly$ that can be freely applied
          to existing types (including type variables). The $\readonly$ operator turns a type into an immutable version of the same type.
          While this complicates the definition of subtyping and proofs of canonical forms lemmas,
          we resolve these issues by reducing types to a normal form.
\end{itemize}

Our hope is to enable reference immutability systems in functional languages by giving simple, sound foundations in \Fsub, a calculus that underpins many practical functional programming languages.

The rest of this paper is organized as follows.  In Section~\ref{sec:reference} we give an overview of reference immutability.  In Section~\ref{sec:dynamic} we introduce an un-typed core calculus, \Lm, to describe {\tt seal}ing
and how it relates to reference immutability safety at run time.
In Section~\ref{sec:metatheory} we present \Fm, which enriches \Lm with types, and show that it satisfies the standard
soundness theorems.  In Section~\ref{section:immutable} we use the soundness results from \Fm and the dynamic safety results from \Lm to show that our desired immutability safety properties hold in \Fm. We survey related and possible future work in Section~\ref{sec:related} and we conclude in Section~\ref{sec:conclusion}.

Our development is mechanized in the Coq artifact that we will submit to the OOPSLA artifact evaluation process.

\section{Reference Immutability}
\label{sec:reference}
Reference immutability at its core is concerned with two key ideas:
\begin{itemize}
    \item {\bf Immutable references:} References to values can be made immutable,
        so that the underlying value {\bf cannot} be modified through that reference.
    \item {\bf Transitive immutability:} An immutable reference to a {\it compound value} that contains other
        references cannot be used to obtain a mutable reference to another value.
        For example, if {\tt x} is a read-only reference to a pair, the result of evaluating {\tt x.first}
        should be {\it viewpoint adapted \cite{10.1007/978-3-540-73589-2_3}} to be a read-only reference, even if the pair contains references that are otherwise mutable.
\end{itemize}

For example, consider the following snippet of Scala-like code that deals with polymorphic mutable pairs.

\begin{lstlisting}[language=Scala]
    case class Pair[X](var first: X, var second: X)

    def good(x : Pair[Int]) = { x.first = 5 }
    def bad1(y : @readonly Pair[Int]) = { y.first = 7 }
    def bad2(y : @readonly Pair[Pair[Int]]) = { y.first.first = 5 }
    def access(z: @readonly Pair[Pair[Int]]): @readonly Pair[Int] = { z.first }
\end{lstlisting}

A reference immutability system would deem the function {\tt good} to be well-typed because it mutates
the pair through a mutable reference {\tt x}. 
However, it would disallow {\tt bad1} because it mutates the pair through a read-only reference {\tt y}.  Moreover, it would also disallow {\tt bad2} because it mutates the pair
referenced indirectly through the read-only reference {\tt y}.  This can also be seen by looking
at the {\tt access} function, which returns a read-only reference of type {\tt @readonly Pair[Int]}
to the first component of the pair referenced by {\tt z}.

\subsection{Why though?}
Immutable values are crucial even in impure functional programming languages because pure code is often easier to reason about. 
This benefits both the programmer writing the code, 
making debugging easier, and the compiler when applying optimizations.

Although most values, even in impure languages, are immutable by default \cite{Haller_2017}, 
mutable values are sometimes necessary for various reasons.  For example, consider a compiler
for a pure, functional, language.  Such a compiler might be split into multiple passes,
one which first builds and generates a {\it symbol table} of {\it procedures} during semantic analysis,
and one which then uses that symbol table during {\it code generation}.  
For efficiency, we may wish to build both the table and the procedures in that table with an impure loop.

\begin{lstlisting}[language=Scala]
object analysis {
    class Procedure(name : String) {
        val locals : mutable.Map[String, Procedure] = mutable.Map.empty
        def addLocalProcedure(name: String, proc: Procedure) = {
            local += (name -> proc)
        }
    }
    
    val table : mutable.Map[String, Procedure] = mutable.Map.empty

    val analyze(ast: AST) = {
        ast.forEach(() => { table.add(new Procedure(...)) })
    }
}
\end{lstlisting}

The symbol table and the properties of the procedure should not be mutable everywhere, though;
during code generation, our compiler should be able to use the information in the
table to generate code but shouldn't be able to change the table nor the information in it!
How do we enforce this though?

One solution is to create an immutable copy of the symbol table for the code generator, but this
can be fragile.  A naive solution which merely clones the table itself will not suffice, for example:
\begin{lstlisting}[language=Scala]
object analysis {
    private val table[analysis] = ...
    def symbolTable : Map[String, Procedure] = table.toMap // create immutable copy of table.
}

object codegen {
    def go() = {
        analysis.symbolTable["main"].locals += ("bad" -> ...) // whoops...
    }
}
\end{lstlisting}

While this does create an immutable copy of the symbol table for the code generator,
it does not create immutable copies of the procedures held in the table itself!
We would need to recursively rebuild a new, immutable symbol table with new, immutable procedures to guarantee immutability, which can be an expensive proposition, both in terms of code and in terms of runtime costs.

Moreover, creating an immutable copy might not even work in all cases.
Consider an interpreter for a pure, functional language with support for {\tt letrec x := e in f}.  
The environment in which  $e$ is interpreted contains a cyclic reference to $x$, which necessitates mutation in the interpreter.  Without special tricks like lazyness this sort of structure cannot be constructed, let alone
copied, without mutation.

\begin{lstlisting}[language=Scala]
abstract class Value { }
type Env = Map[String, Value]
case class Closure(var env: Env, params: List[String], body: Exp) extends Value

def interpret_letrec(env: Env, x: String, e: Exp, f: Exp) : Value = {
    val v = interpret(env + (x -> Nothing), e)
    case v of {
        Closure(env, params, body) => v.env = v.env + (x -> v) // Update binding
    }
    interpret (env + (x -> v), f)
}
\end{lstlisting}

Here, the closure that {\tt v} refers to needs to be mutable while it is being constructed,
but since the underlying language is pure, it should be immutable afterwards.  In particular,
we should not be able to mutate the closure through the self-referential reference
{\tt v.env = env + (x -> v)}, nor should we be able to mutate the closure while interpreting {\tt f}.

We would like a system that prevents writes to {\tt v} from the self-referential
binding in its environment and from the reference we pass to {\tt interpret (env + (x -> v), f)}.
This is what {\it reference immutability} provides.

\begin{lstlisting}[language=Scala]
abstract class Value { }
type Env = Map[String, @readonly Value]
case class Closure(env: var Env, params: List[String], body: Exp)

def interpret_letrec(env: Env, x: String, e: Exp, f: Exp) : Value = {
    val v = interpret(env + (x -> Nothing), e)
    case v of {
        Closure(env, params, body) => v.env = env + (x -> @readonly v) // update binding
    }
    interpret (env + (x -> @readonly v), f)
}

\end{lstlisting}

\section{Dynamic Immutability Safety}
\label{sec:dynamic}
Now, to formalize reference immutability, we need to formalize exactly when references are used
to update the values they refer to.  For example, from above, how do we check that {\tt access}
does what it {\it claims} to do?
\begin{lstlisting}[language=Scala]
    def access(z: @readonly Pair[Pair[Int]]): @readonly Pair[Int] = { z.first }
\end{lstlisting}

How do we check that {\tt access} returns a reference
to {\tt z.first} that, at runtime, is never used to write to {\tt z.first} or any other values transitively reachable from it through other references? How do we even express this guarantee precisely?

If we consider a reference as a collection of getter and setter methods for the fields of the
object it refers to, we could ensure that a reference is immutable by dropping all the setter
methods. To ensure that immutability is transitive, we would also need to ensure that the result
of applying a getter method is also immutable, i.e. by also dropping its setter methods and recursively
applying the same modification to \emph{its} getter methods. We will make this precise by introducing
the \Lm calculus with a notion of \emph{sealed} references.

\subsection{\Lm}
To answer this question we introduce \Lm, the untyped lambda calculus with collections of mutable references -- namely, records
-- extended with a mechanism for {\it sealing} references.   \Lm is adapted from the CS-machine of
\citet{10.1145/41625.41654} and extended with rules for dealing with sealed references.

\paragraph{\textbf{Sealed references:}} To address the question about dynamic, runtime safety -- can we ensure that read-only references are never used to mutate values -- references can be explicitly {\it sealed} so
that any operation that will mutate the cell referenced will fail to evaluate; see Figure~\ref{fig:Lm}.

\begin{figure}
  \begin{minipage}{0.45\textwidth}
  \[
  \begin{array}[t]{rll@{\hspace{4mm}}l}\\
   s, t & ::= & & \mbox{\bf\textsf{Terms}} \\
        & | & \lambda x . t & \mbox{term abstraction} \\
        & | & x & \mbox{term variable} \\
        & | & s(t) & \mbox{application} \\
        & | & \{f_1 : s_1, f_2 : s_2, \hdots\} & \mbox{records} \\
        & | & s.f & \mbox{field read} \\
        & | & s.f = t & \mbox{field write} \\
        & | & \highlight{\seal s} & \mbox{sealing}
  \end{array}
  \]
  \end{minipage}\hfill
  \begin{minipage}{0.5\textwidth}
    \[
      \begin{array}[t]{rll@{\hspace{4mm}}l}
             &  & l & \mbox{\bf\textsf{Location}} \\[\medskipamount]
       s, t & ::= & & \mbox{\bf\textsf{Runtime Terms}} \\
            & |   & \{x_1 : l_1, x_2 : l_2, \hdots\} & \mbox{runtime record}\\[\medskipamount]
                  
      v   & ::= & & \mbox{\bf\textsf{Runtime Values}} \\
          & | & \lambda x . t & \\
          & | & \{f_1 : l_1, f_2 : l_2, \hdots\} \\
          & | & \highlight{\seal \{f_1, l_1, f_2 : l_2, \hdots\}} \\
      \end{array}
    \]
    \end{minipage}

  \begin{minipage}{0.45\textwidth}{\small
    \vspace{1.1em}
    \infax[\ruledefN{beta-v}{beta-v}]{
        \langle (\lambda x.t)(v), \sigma \rangle \reduces \langle t[x \mapsto v], \sigma \rangle
    }
    \vspace{1em}
    \infrule[\ruledefN{record-store}{record-store}]{
        l_i \notin \sigma
    }{
        \langle \{x_i : v_i\}, \sigma \rangle \reduces \langle \{x_i : l_i\}, (\sigma, l_1 : v_1, l_2 : v_2, \hdots) \rangle 
    }
    \vspace{1em}
    \infrule[\ruledefN{field}{field}]{
        l : v \in \sigma 
    }{
        \langle \{\hdots x : l \hdots\}.x, \sigma \rangle \reduces \langle v, \sigma \rangle
    }
    \vspace{1em}
    \vspace{1em}
    \infrule[\ruledefN{write-field}{write-field}]{
        l : v \in \sigma 
    }{
        \langle \{\hdots x : l \hdots\}.x = v', \sigma \rangle \reduces \langle v, \sigma[l \mapsto v'] \rangle
    }}
  \end{minipage}\hfill
    \begin{minipage}{0.45\textwidth}{\small
    
      \infrule[\ruledefN{sealed-field}{sealed-field}]{
          l : v \in \sigma 
      }{
          \langle \highlight{(\seal \{\hdots x : l \hdots\})}.x, \sigma \rangle \reduces \langle \highlight{\seal v}, \sigma \rangle
      }
      \vspace{1em}
      \infax[\ruledefN{seal-elim-abs}{seal-elim-abs}]{
          \langle \highlight{\seal (\lambda x . t)}, \sigma \rangle \reduces \langle \lambda x . t, \sigma \rangle
      }
      \vspace{1em}
      \vspace{1em}
      \infax[\ruledefN{seal-elim-multiple}{seal-elim-multiple}]{
          \langle \highlight{\seal \seal v}, \sigma \rangle \reduces \langle \seal v, \sigma\rangle
      }
      \vspace{1em}
      \infrule[\ruledefN{context}{context}]{
          \langle s, \sigma \rangle \reduces \langle t, \sigma' \rangle
      }{
          \langle E[s], \sigma \rangle \reduces \langle E[t], \sigma' \rangle
      }}  
  \end{minipage}

  \begin{center}
    \[
    \begin{array}{lcll}
        E & ::= & []~|~E(t)~|~v(E) & \mbox{{\bf Evaluation Context}} \\
            & |   & \{x_0 : v_0, \hdots, x_i : E, x_{i+1} : t_{i+1}, \hdots\}\\
            & |   & E.x \\
            & |   & E.x = t~|~v.x = E \\
            & |   & \highlight{\seal E}
    \end{array}
    \]
    \end{center}

  \caption{The syntax and semantics of $\lambda_m$.}
  \label{fig:Lm}

\end{figure}

\newcommand{\addr}[1]{\text{\tt 0x{#1}}}

The {\tt seal} form protects its result from writes.  A term under a {\tt seal} form reduces
until it becomes a value.  At that point, values that are not records,
like functions and type abstractions, are just transparently passed through the {\tt seal} construct.  However, values that {\it are} -- records -- remain protected by the {\tt seal} form, and do not reduce further.  For example:
\[\seal(\{y : \addr{0001}\})\]
is an irreducible value -- a sealed record where the first field is stored at location $1$ in the store.
Intuitively, this can be viewed as removing the setter methods from an object reference.  A sealed
reference $\seal v$ behaves {\it exactly} like its unsealed variant $v$ except that writes to $\seal v$ are forbidden and reads from $\seal v$ return {\tt seal}ed results.

Rules that mutate the cells corresponding to a record 
explicitly require an unsealed open record; see \ruleref{write-field}. 
This ensures that any ill-behaved program that mutates a store cell through a sealed record will get stuck,
while an unsealed record can have its fields updated:
\[
\begin{array}{ccl}
     \langle \{x : 10\}.x = 5, []\rangle & \reduces & \langle \{x : \addr{0001}\}.x = 5, [\addr{0001}: 10] \rangle \\
        & \reduces & \langle 10, [\addr{0001}: 5] \rangle 
\end{array}
\]

A sealed record cannot have its fields written to. Unlike record field reads, for which there is a
sealed \ruleref{sealed-field} counterpart to the standard record read rule \ruleref{field},
there is no corresponding rule for writing to a sealed record for \ruleref{write-field}.
Recall that \ruleref{write-field} requires an {\it open, unsealed} record as input:
    \infrule[]{
        l : v \in \sigma 
    }{
        \langle \{\hdots x : l \hdots\}.x = v', \sigma \rangle \reduces \langle v, \sigma[l \mapsto v'] \rangle
    }
The calculus does not contain any rule like the following, which would reduce writes on a sealed record:
    \infrule[]{
        l : v \in \sigma 
    }{
        \langle \highlight{(\seal \{\hdots x : l \hdots\})}.x = v', \sigma \rangle \reduces \langle v, \sigma[l \mapsto v'] \rangle
    }
So a term like:
\[
\begin{array}{ccl}
     \langle (\seal \{x : 10\}).x = 5, []\rangle & \reduces & \langle \seal(\{x : \addr{0001}\}).x = 5, [\addr{0001}: 10] \rangle \\
        & \reduces & \mbox{gets stuck.}
\end{array}
\]

\paragraph{\textbf{Dynamic viewpoint adaptation:}}  After reading a field from a sealed record, the
semantics {\it seals} that value,
ensuring transitive safety -- see \ruleref{sealed-field}. 
    \infrule[]{
        l : v \in \sigma 
    }{
        \langle \highlight{(\seal \{\hdots x : l \hdots\})}.x, \sigma \rangle \reduces \langle \highlight{\seal v}, \sigma \rangle
    }

For example:
\[
\begin{array}{ccl}
     \langle (\seal \{y : \{x : 10\}\}).y, []\rangle & \reduces & \langle \seal(\{y : \{x: \addr{001}\}\}).y, [\addr{001}: 10] \rangle \\
        & \reduces & \langle \seal(\{y : \addr{002}\}).y, [\addr{001}: 10, \addr{002}: \{x: \addr{001}\}] \rangle \\
        & \reduces & \langle \seal (\{x: \addr{001}\}), [\addr{001}: 10, \addr{002}: \{x: \addr{001}\}] \rangle
\end{array}
\]

{\it Sealed references} and {\it dynamic viewpoint adaptation} allow for a succinct guarantee
of {\it dynamic transitive immutability safety} -- that no value is ever mutated through a read-only
reference or any other references transitively derived from it.

Aside from preventing writes through sealed references, we should show that
sealing does not otherwise affect reduction.
For this we need a definition that
relates pairs of terms that are essentially equivalent except that one has more seals than the other.
\begin{definition}
    Let $s$ and $t$ be two terms.  We say $s \leq t$ if $t$ can be obtained from $s$
    by repeatedly replacing sub-terms $s'$ of $s$ with sealed subterms $\seal s'$.
\end{definition}

This implies a similar definition for stores:
\begin{definition}
    Let $\sigma$ and $\sigma'$ be two stores.  We say $\sigma \leq \sigma'$ if and only if
    they have the same locations and for every location $l \in \sigma$, we have $\sigma(l) \leq \sigma'(l)$.
\end{definition}

The following three lemmas formalize how reduction behaves for terms that are equivalent modulo seals.
The first one is for a term $t$ that is equivalent to a value -- it states that if $t$
reduces, the resulting term is still equivalent to the same value.
It also shows that the resulting term has fewer seals than $t$, which we'll need
later for an inductive argument.
\begin{definition}
    Let $s$ be a term.  Then $|s|$ is the number of seals in s.
\end{definition}
\begin{lemma}
    \label{lemma:un-safety-value}
    Let $v$ be a value, $\sigma_v$ be a store, $t$ be a term such that $v \leq t$, and $\sigma_t$ be
    a store such that $\sigma_v \leq \sigma_t$.
    
    If $\langle t, \sigma_t \rangle \reduces \langle t', \sigma_t'\rangle$
    then $v \leq t'$, $\sigma_v \leq \sigma_t'$, and $|t'| < |t|$. 
\end{lemma}

The next lemma is an analogue of Lemma~\ref{lemma:un-safety-value} for terms.  Given two equivalent
terms $s$ and $t$, if $s$ steps to $s'$ and $t$ steps to $t'$, then
either $s$ and $t'$ are equivalent or $s'$ and $t'$ are equivalent.  Moreover, again,
to show that reduction in $t$ is equivalent to reduction in $s$, we have that $|t'| < t$ if $s \leq t'$.
\begin{lemma}
    \label{lemma:un-safety-step}
    Let $s, t$ be terms such that $s \leq t$ and let $\sigma_s, \sigma_t$ be stores such that
    $\sigma_s \leq \sigma_t$.  If $\langle s, \sigma_s\rangle \reduces \langle s', \sigma_s'\rangle$
    and $\langle t, \sigma_t\rangle \reduces \langle t', \sigma_t'\rangle$ then:
    \begin{enumerate}
        \item Either $s \leq t'$, $\sigma_s \leq \sigma_t'$, and $|t'| < |t|$, or
        \item $s' \leq t'$ and $\sigma_s' \leq \sigma_t'$.
    \end{enumerate}
\end{lemma}

Together, Lemmas \ref{lemma:un-safety-value} and \ref{lemma:un-safety-step} relate how terms
$s$ and $t$ reduce when they are equivalent modulo seals.  Assuming that both $s$ and $t$ reduce, every step of $s$ corresponds to  finitely many steps of $t$, and they reduce to equivalent results as well.
This shows that sealing is transparent when added onto references that are never written to,
allowing for a succinct guarantee of immutability safety.

Finally, the last lemma states that erasing {\tt seal}s will never cause a term to get stuck.
Seals can be safely erased without affecting reduction.
\begin{lemma}
    \label{lemma:un-safety-step-exists}
    Let $s, t$ be terms such that $s \leq t$ and let $\sigma_s, \sigma_t$ be stores such that
    $\sigma_s \leq \sigma_t$.  If $\langle t, \sigma_t\rangle \reduces \langle t', \sigma_t'\rangle$ then:
    \begin{enumerate}
        \item Either $s \leq t'$, $\sigma_s \leq \sigma_t'$, and $|t'| < |t|$, or
        \item There exists $s'$ and $\sigma_s'$ such that $\langle s, \sigma_s\rangle \reduces \langle s', \sigma_s'\rangle$, $s' \leq t'$ and $\sigma_s' \leq \sigma_t'$.
    \end{enumerate}
\end{lemma}
\newcommand{\rreduces}{\;\longrightarrow^*\;}

From this we can derive the following multi-step analogue, after observing the following lemma:
\begin{lemma}
    \label{lemma:seal-value}
    If $s$ is a term and $v$ is a value such that $s \leq v$, then $s$ is also a value.
\end{lemma}

Hence:
\begin{lemma}
    \label{lemma:un-erasure}
    Suppose $s$ and $t$ are terms such that $s \leq t$. If $\langle t, \sigma_t\rangle \rreduces \langle v_t, \sigma_t'\rangle$ for some value $v_t$, then for any $\sigma_s \leq \sigma_t$
    we have $\langle s, \sigma_s\rangle \rreduces \langle v_s, \sigma_s'\rangle$ such that
    $v_s' \leq v_s'$ and $\sigma_s' \leq \sigma_t'$.
\end{lemma}

Finally, it can be shown that the seals are to blame when two equivalent terms $s$ and $t$ reduce differently
-- in particular, when one reduces but the other gets stuck.
\begin{lemma}
    \label{lemma:write-field-stuck}
    Let $s, t$ be terms such that $s \leq t$, and let $\sigma_s, \sigma_t$ be stores such that
    $\sigma_s \leq \sigma_t$.  If $\langle s, \sigma_s\rangle \reduces \langle s', \sigma_s'\rangle$
    and $t$ gets stuck, then the reduction performed on $s$ was a write to a record using rule \ruleref{write-field}. 
\end{lemma}
\begin{proof}
    {\it (Sketch)} As $s$ cannot further reduce, the evaluation context of $s$ and $t$ must match; there
    are no extraneous {\tt seal}s that need to be discharged.  As such, from inspection of the reduction rules,
    we see that in all cases except for \ruleref{write-field}, for every possible reduction that $s$
    could have taken, there is a possible reduction that $t$ could have taken as well, as desired.
\end{proof}

\section{Typing and Static Safety}
\label{sec:metatheory}

\Lm provides a {\it dynamic guarantee}
that a given program will never modify its sealed references, but it does not provide any
static guarantees about the dynamic behavior of a given program. To do that, we need a type
system for \Lm that will reject programs 
like {\tt access(seal Pair(3,5)).first = 10},
which we know will crash.

To ensure that well-typed programs do not get stuck, a type system for \Lm needs a static analogue of {\tt seal}ing -- a way to turn an existing type into a {\it read-only type}. 
Read-only types denote references that are {\it immutable} and that (transitively) {\it adapt}
any other references read through them to be {\it immutable} as well.

Issues arise, however, when we introduce polymorphism.

\subsection{Polymorphism}
Recall our earlier example -- a polymorphic {\tt Pair} object.
\begin{lstlisting}[language=Scala]
    case class Pair[X](var first: X, var second: X)
\end{lstlisting}

In a functional language, it is only natural to write higher-order functions that are polymorphic
over the elements stored in the pair.  Consider an in-place map function over pairs, which applies a function
to each element in the pair, storing the result in the original pair.  This naturally requires mutable
access to a pair.

\begin{lstlisting}[language=Scala]
    def inplace_map[X](pair: Pair[X], f: X => X): Unit = {
        pair.first = f(pair.first);
        pair.second = f(pair.second);
    }
\end{lstlisting}

This is all well and good, but we may wish to restrict the behaviour of {\tt f} over the elements of the
pair.  It may be safer to restrict the behaviour of {\tt f} so that it could not mutate the elements
passed to it.  Note that we cannot restrict access to the pair, however, as we still need to mutate it.

\begin{lstlisting}[language=Scala]
    // Is this well founded?
    def inplace_map[X](pair: Pair[X], f: @readonly X => X): Unit = {
        pair.first = f(pair.first);
        pair.second = f(pair.second);
    }
\end{lstlisting}

Now, such a definition requires the ability to further modify type variables with immutability
qualifiers.  This raises important questions -- for example, is this operation even well founded?  This
depends on what {\tt X} ranges over.

\paragraph*{X ranges over an unqualified type:}  If type variables range over types which
have not been qualified by {\tt @readonly}, then
this operation is clearly well founded -- it is simply qualifying the unqualified type that
{\tt X} will eventually be substituted by with the {\tt @readonly} qualifier.  This approach
has been used by ReIm for Java and for an immutability system for C\# -- \cite{10.1145/2384616.2384680,10.1145/2384616.2384619}.

However, this raises the problem of polymorphism over immutability
qualifiers as well -- for example, a {\tt Pair} should be able to store both immutable and mutable object
references.  The only natural solution is to then introduce a {\it mutablity} qualifier binder
to allow for polymorphism over immutability qualifiers, as thus:
\begin{lstlisting}[language=Scala]
    case class Pair[M, X](var first: M X, var second: M X)
    def inplace_map[M, X](pair: Pair[M, X], f: @readonly X => M X): Unit = {
        pair.first = f(pair.first);
        pair.second = f(pair.second);
    }
\end{lstlisting}

Mutability qualifier binders have been used previously, most notably by \cite{10.1145/2384616.2384619}.
For one, updating the binding structure of a language is not an easy task -- ReIm notably {\it omits}
this sort of parametric mutability polymorphism \cite{10.1145/2384616.2384680}.
However, this sort of solution has its downsides; in particular, existing higher-order functions need to be updated with immutability annotations or variables, as type variables no longer stand for a full type.
For example, an existing definition of List {\tt map} which appears as thus originally:
\begin{lstlisting}[language=Scala]
    def map[X](l: List[X], f: X => X): List[X]
\end{lstlisting}
needs to be updated to read as the following instead: 
\begin{lstlisting}[language=Scala]
    def map[M, X](l: List[M X], f: M X => M X): List[M X]
\end{lstlisting}

Instead, we would like to have {\tt X} range over fully qualified types as well, but as we will
see that poses some issues as well.

\paragraph{X ranges over fully-qualified types:}  If type variables can range over
types which have been already qualified by {\tt @readonly}, then we can avoid introducing mutability binders in the definitions for {\tt Pair}, {\tt inplace\_map},
and {\tt map} above.  A {\tt Pair} can be polymorphic over its contents {\tt X} without caring about
the underlying mutability of {\tt X}.  However, this raises the question -- how do we interpret repeated
applications of the {\tt @readonly} qualifier?  For example, what if we applied {\tt inplace\_map} on a
{\tt Pair[@readonly Pair[Int]]}?  Then {\tt inplace\_map} would expect a function {\tt f} with type 
{\tt @readonly (@readonly Pair[Int]) => @readonly Pair[Int]}.  While our intuition would tell us that {\tt @readonly (@readonly Pair[Int])} is really just a {\tt @readonly Pair[Int]}, discharging this equivalence in a proof is not so easy.

One response is to explicitly prevent type variables from being further qualified.  Calculi which take this
approach include  \cite{10.1145/1287624.1287637, 10.1145/1103845.1094828}.  However,
this restriction prevents this version of {\tt inplace\_map} from being expressed. How can we address this?

Our approach, which we explain below, is to treat {\tt @readonly} as a type operator that works over all types.  
Following the intuition that {\tt seal}ing removes setters from references, {\tt @readonly} should be a type operator which removes setters
from types.  While this does cause complications, we show below how types like {\tt @readonly @readonly Pair[Int]}
can be dealt with, using {\it subtyping} and {\it type normalization}.

\subsection{\Fm}

To address these issues, we introduce \Fm, which adds a type system in the style of \Fsub to \Lm.  The syntax of \Fm is given in Figure \ref{fig:syntax};
changes from \Fsub are noted in grey.

\begin{figure}
  \begin{minipage}{0.47\textwidth}
  \[
  \begin{array}[t]{rll@{\hspace{4mm}}l}\\
   s, t & ::= & & \mbox{\bf\textsf{Terms}} \\
        & | & \lambda x . t & \mbox{term abstraction} \\
        & | & \Lambda (X\sub S) . t & \mbox{type abstraction} \\
        & | & x & \mbox{term variable} \\
        & | & s(t) & \mbox{application} \\
        & | & s[T] & \mbox{type application} \\
        & | & \{f_1 : s_1, f_2 : s_2, \hdots\} & \mbox{records} \\
        & | & s.x & \mbox{field read} \\
        & | & s.x = t & \mbox{field write} \\
        & | & \highlight{\seal s} & \mbox{sealing}
  \end{array}
  \]
  \end{minipage}\hfill
  \begin{minipage}{0.47\textwidth}
  \[
  \begin{array}[t]{rll@{\hspace{4mm}}l}\\
   S, T & ::= & & \mbox{{\bf\textsf{Types}}} \\
          & | & X & \mbox{type variable} \\
          & | & S \to T & \mbox{function type} \\
          & | & \forall(X\texttt{<:}S).T & \mbox{for-all type} \\
          & | & S \wedge T & \mbox{intersection type} \\
          & | & \{f : T\} & \mbox{record type} \\
          & | &  \highlight{\readonly T} & \mbox{readonly type} \\[\medskipamount]

  \Gamma & ::= & & \mbox{\bf\textsf{Environment}} \\
         & | & \cdot & \mbox{empty} \\
         & | & \Gamma,~x : T & \mbox{term binding} \\
         & | & \Gamma,~X <: T & \mbox{type binding} \\
  \end{array}
  \]
  \end{minipage}
  
  \begin{minipage}{0.47\textwidth}
  \[
    \begin{array}[t]{rll@{\hspace{4mm}}l}
           &  & l & \mbox{\bf\textsf{Location}} \\[\medskipamount]
     s, t & ::= & & \mbox{\bf\textsf{Runtime Terms}} \\
          & |   & \{x_1 : l_1, x_2 : l_2, \hdots\} & \mbox{runtime record}\\[\medskipamount]
                
    v   & ::= & & \mbox{\bf\textsf{Runtime Values}} \\
        & | & \lambda x . t & \\
        & | & \Lambda (X \sub S) . t \\
        & | & \{f_1: l_1, f_2 : l_2, \hdots\} \\
        & | & \highlight{\seal \{f_1, l_1, f_2 : l_2, \hdots\}} \\
    \end{array}
  \]
  \end{minipage}\hfill
  \begin{minipage}[c]{0.47\textwidth}
  \[
    \begin{array}[t]{rll@{\hspace{4mm}}l} \\[\medskipamount]
     \sigma & ::= & & \mbox{\bf\textsf{Store}} \\
            & |   & \cdot & \mbox{empty} \\
            & |   & \sigma,~l : v & \mbox{cell $l$ with value $v$} \\[\medskipamount]

     \Sigma & ::= & & \mbox{\bf\textsf{Store Environment}} \\
            & |   & \cdot & \mbox{empty} \\
            & |   & \sigma,~l : T & \mbox{cell binding} \\[\medskipamount]
            \\[\medskipamount]
            \\[\medskipamount]
            \\[\medskipamount]
    \end{array}
  \]
  \end{minipage}
  \caption{The syntax of \Fm.}
  \label{fig:syntax}

\end{figure}

\Fm is a straightforward extension of \Fsub with collections of mutable references -- namely, records
-- and with two new extensions: {\it read-only} types and {\it sealed} references.  To be close to existing
functional languages with subtyping and records, records in \Fm are modelled
as intersections of single-element record types, to support record subsumption, as in \cite{amin2016essence} and
\cite{Reynolds1997}.
See Figures~\ref{fig:subtyping} and~\ref{fig:typing} for full subtyping and typing rules respectively.

\begin{figure}
  \judgement{Normal Forms}{}
  \[
  \begin{array}[t]{rll@{\hspace{4mm}}l}\\
   S, T & ::= & \mbox{\bf\textsf{Types in normal form}} \\
        & |   & \bigwedge_i (R_i)  & \mbox{Intersection of components} \\
   R    & ::= & \mbox{\bf\textsf{Normal form type components}} \\
        & |   & \top & \mbox{Top type} \\
        & |   & S \to T & \mbox{Normal function type} \\
        & |   & \forall(X\sub S).T & \mbox{Normal for-all type} \\
        & |   & \{f : S\} & \mbox{Normal record type} \\
        & |   & X & \mbox{Type variable} \\
        & |   & \readonly \{f : S\} & \mbox{Read-only normal record type} \\
        & |   & \readonly X & \mbox{Read-only type variable} \\
  \end{array}
  \]

  \caption{Normal forms for \Fm.}
\label{fig:normalization}

\end{figure}

\begin{figure}
  \judgement{Subtyping}{\fbox{$\Gamma \vdash S \sub T$}}

  \begin{minipage}{0.45\textwidth}
  \vspace{1.1em} 
  \infax[\ruledef{refl}]{%
    \Gamma \vdash T \sub T}

  \vspace{1em}

  \infrule[\ruledef{trans}]{%
    \Gamma \vdash R \sub S \gap \Gamma \vdash S \sub T
  }{%
    \Gamma \vdash R \sub T}

  \vspace{1em}

  \infrule[\ruledef{tvar}]{%
    X \sub T \in \Gamma
  }{%
    \Gamma \vdash X \sub T}

  \vspace{1em}

  \infax[\ruledef{top}]{%
    \Gamma \vdash U \sub \top}
  \vspace{1em}

  \infrule[\ruledef{arrow}]{
    \Gamma \vdash T_1 \sub S_1 \gap \Gamma \vdash S_2 \sub T_2   
  }{
    \Gamma \vdash S_1 \to S_2 \sub T_1 \to T_2
  }

\vspace{1em}

  \infrule[\ruledef{all}]{
    \Gamma \vdash T_1 \sub S_1 \gap \Gamma, X \sub T_1 \vdash S_2 \sub T_2
  }{
    \Gamma \vdash \forall(X \sub S_1).S_2 \sub \forall(X \sub T_1).T_2
  }
  \end{minipage}\hfill
  \begin{minipage}{0.45\textwidth}
  \vspace{1.1em}



    \infrule[\ruledef{record}]{
        \Gamma \vdash S \sub T \gap \Gamma \vdash T \sub S
    }{
        \Gamma \vdash \{x : S\} \sub \{x : T\}
    }
    
    \vspace{1em}

    \infrule[\ruledef{readonly-record}]{
        \Gamma \vdash S \sub T
    }{
        \highlight{\Gamma \vdash \readonly \{x : S\} \sub \readonly \{x : T\}}
    }    

    \vspace{1em}

    \infax[\ruledef{inter-left}]{
        \Gamma \vdash S \wedge T \sub S
    }

    \vspace{1em}

    \infax[\ruledef{inter-right}]{
        \Gamma \vdash S \wedge T \sub T
    }
    \vspace{1em}

    \infrule[\ruledef{inter}]{
        \Gamma \vdash S \sub T_1 \gap \Gamma \vdash S \sub T_2 
    }{
        \Gamma \vdash S \sub T_1 \wedge T_2
    }
    \vspace{1em}

  \end{minipage}

    \begin{center}
    \begin{minipage}{0.75\textwidth}
    \vspace{1.1em}
        \infrule[\ruledef{readonly}]{
            \Gamma \vdash S \sub T
        }{
            \highlight{\Gamma \vdash \readonly S \sub \readonly T}
        }

        \vspace{1em}
        \infrule[\ruledef{mutable}]{
            \Gamma \vdash S \sub T
        }{
            \highlight{\Gamma \vdash S \sub \readonly T}
        }
        
        \vspace{1em}
        \infrule[\ruledef{denormalize}]{
            \Gamma \vdash nf(S) \sub nf(T)
        }{
            \highlight{\Gamma \vdash S \sub  T}
        }
    \end{minipage}
    \end{center}
\caption{Subtyping rules of \Fm.}
\label{fig:subtyping}

\end{figure}

\begin{figure}
  \judgement{Typing and Runtime Typing}{\fbox{$\Gamma~|~\highlight{\Sigma} \vdash t : T$ and $\Gamma~|~\highlight{\Sigma} \vdash \highlight{\sigma}$}}

  \begin{minipage}{0.45\textwidth}
  \vspace{1.1em} 
  \infrule[\ruledef{var}]{
    x : T \in \Gamma
  }{
    \Gamma~|~\Sigma \vdash x : T
  }
  \vspace{1em}
  \infrule[\ruledef{abs}]{
    \Gamma, x : S \vdash t : T  
  }{
    \Gamma~|~\Sigma \vdash \lambda x . t : S \to T
  }
  
  \vspace{1em}
  \infrule[\ruledef{t-abs}]{
    \Gamma, X \sub S \vdash t : T
  }{
    \Gamma~|~\Sigma \vdash \Lambda (X \sub S).t : \forall(X \sub S).T
  }
  
  \vspace{1em}
  \infrule[\ruledef{app}]{
    \Gamma~|~\Sigma \vdash t : S \to T \gap \Gamma~|~\Sigma \vdash s : S 
  }{
    \Gamma~|~\Sigma \vdash t(s) : T
  }
  
  \vspace{1em}
  \infrule[\ruledef{t-app}]{
    \Gamma~|~\Sigma \vdash t : \forall(X \sub S).T \gap \Gamma~|~\Sigma \vdash S' \sub S 
  }{
    \Gamma~|~\Sigma \vdash t[S'] : T[X \mapsto S']
  }

  \end{minipage}\hfill
  \begin{minipage}{0.45\textwidth}

  \vspace{1.1em} 
  
  
  
  \infrule[\ruledef{record-intro}]{
    \Gamma~|~\Sigma \vdash t_i : T_i
  }{
    \Gamma~|~\Sigma \vdash \{x_i : t_i \hdots\} : \bigwedge_{i} \{x_i : T_i\}
  }
  
  \vspace{1em}
  \infrule[\ruledef{record-elim}]{
    \Gamma~|~\Sigma \vdash t : \{x : T\}
  }{
    \Gamma~|~\Sigma \vdash t.x : T
  }

  \vspace{1em}
  \infrule[\ruledef{record-update}]{
    \Gamma~|~\Sigma \vdash s : \{x : T\} \gap \Gamma \vdash t : T
  }{
    \Gamma~|~\Sigma \vdash s.x = t : T
  }

\vspace{1em}
  \infrule[\ruledef{sub}]{
    \Gamma~|~\Sigma \vdash s : S \gap \Gamma \vdash S \sub T
  }{
    \Gamma~|~\Sigma \vdash s : T
  }
  \end{minipage}

    \vspace{1em}
    \begin{minipage}{0.4\textwidth}
    \infrule[\ruledef{seal}]{
        \Gamma~|~\Sigma \vdash s : S 
    }{
        \highlight{\Gamma~|~\Sigma \vdash \seal s : \readonly S}
    }
    \vspace{1em}
    \infrule[\ruledef{readonly-record-elim}]{
        \Gamma~|~\Sigma \vdash s : \readonly \{x : S\}
    }{
        \highlight{\Gamma~|~\Sigma \vdash s.x : \readonly S}
    }
    \end{minipage}\hfill
    \begin{minipage}{0.45\textwidth}
    \infrule[\ruledef{runtime-record}]{
        l_i : T_i \in \highlight{\Sigma}
    }{
        \highlight{\Gamma~|~\Sigma \vdash \{x_i : l_i\}: \bigwedge_i \{x_i : T_i\}}
    }
    \end{minipage}

    \begin{center}
      \begin{minipage}{0.75\textwidth}
      \infrule[\ruledef{store}]{
        dom(\sigma) = dom(\Sigma) \gap \forall l \in dom(\Sigma),~\Gamma~|~\Sigma \vdash \sigma(l) : \Sigma(l)
      }{
        \Gamma~|~\Sigma \vdash \highlight{\sigma}
      }
      \end{minipage}
    \end{center}
    \caption{Typing rules for \Fm}
    \label{fig:typing}

\end{figure}

\paragraph{\textbf{Read-only types:}} 
The {\tt readonly} type operator transforms an existing type to a read-only version of itself.
Unlike the read-only mutability qualifier in Javari and ReIm, which is paired with a type
to form a pair of a qualifier and a type, a read-only type in \Fm is itself a type.
The {\tt readonly} operator can be seen as the static counterpart of sealing or of deleting setter methods from an object-oriented class type.

Any type {\tt T} is naturally a subtype of its readonly counterpart {\tt \readonly T}, which
motivates the choice of \Fsub as a base calculus. This subtyping relationship
is reflected in the subtyping rule \ruleref{mutable}. The \ruleref{seal} typing rule gives a read-only type to sealed references.

\paragraph{\textbf{Static viewpoint adaptation:}}  The \ruleref{readonly-record-elim} rule is a static counterpart
of the \ruleref{sealed-field} reduction rule. Given a reference $s$ to a record with read-only type, it gives a
read-only type to the result of a read $s.x$ of a field $x$ from that reference. If $S$ is the type of field
$x$ in the record type given to $s$, the rule viewpoint-adapts the type, giving $s.x$ the type $\readonly S$.

\subsubsection{Normal Forms for Types}
\label{subsection:normal}
In \Fm, \readonly is a type operator that can be applied to any type, which
enables us to express types such as $\readonly X$, where $X$ is some type
variable of unknown mutability. However, if $X$ is itself instantiated with
some readonly type $\readonly T$, the type $\readonly X$ becomes
$\readonly \readonly T$, with two occurrences of the type operator.
Intuitively, such a type should have the same meaning as $\readonly T$.

Additionally, certain types should be equivalent under subtyping.  For example, 
for both backwards compatibility and simplicity, arrow $S \to T$ and for-all types $\forall (X\sub S).T$ should be equivalent under subtyping to their read-only forms $\readonly (S \to T)$ and $\readonly (\forall (X\sub S).T)$, respectively, as well.

Having multiple representations for the same type, even infinitely many,
complicates reasoning about the meanings of types and proofs of soundness.
Therefore, we define a canonical representation for types as follows:

\begin{definition}
    A type $T$ is in normal form if:
    \begin{enumerate}
        \item $T$ is the top type $\top$.
        \item $T$ is a function type $S_1 \to S_2$, where $S_1$ and $S_2$
                are in normal form.
        \item $T$ is an abstraction type $\forall(X \sub S_1).S_2$, where $S_1$
                and $S_2$ are in normal form.
        \item $T$ is an intersection type $S_1 \wedge S_2$, where $S_1$ and $S_2$
                are in normal form.
        \item $T$ is a record type $\{ x : S \}$, where $S$ is in normal form.
        \item $T$ is a read-only record type $\readonly \{ x : S \}$, where $S$
                is in normal form.
        \item Type variables $X$ and read-only type variables $\readonly X$
                are in normal form.
    \end{enumerate}
\end{definition}

A type in normal form is simple -- it is an intersection of function, abstraction,
and record types, each possibly modified by a single {\tt readonly} operator.  
For example, $\{x : X\} \wedge \readonly \{y : Y\}$ is in normal form.  The type $\readonly (\{x : X\} \wedge \{y : Y\})$
is not.  A grammar for types in normal form can be found in Figure~\ref{fig:normalization}.

This allows us to reason
about both the shape of the underlying value being typed, and whether or not it has been modified by
a {\tt readonly} operator.  Naturally we need a theorem which states that every type has a normal form and
a function $nf$ to compute that normal form. Such a function $nf$ is shown in Figure \ref{fig:normalizationfn}.
Normalization both computes a normal form and is idempotent -- a type in normal form normalizes to itself.

\begin{figure}
  \judgement{Normalization}{\fbox{$nf(T)$ and $merge(T)$}}

  \[
  \begin{array}[t]{rlll}\\
   nf(T) & ::= & &\mbox{\bf\textsf{Normalization}} \\
         & |~\top & => & \top \\
         & |~X & => & X \\
         & |~S \to T & => & nf(S) \to nf(T) \\
         & |~\forall(X \sub S).T & => & \forall(X \sub nf(S)).nf(T) \\
         & |~S \wedge T & => & nf(S) \wedge nf(T) \\
         & |~\{f : T\} & => & \{f : nf(T)\} \\
         & |~\readonly T & => & merge(T) \\
  \end{array}
  \]

 \[
   \begin{array}[t]{rlll}\\
   merge(T) & ::= & &\mbox{\bf\textsf{Merging}} \\
         & |~X & => & \readonly X \\
         & |~\{f : T\} & => & \readonly \{f : T\} \\
         & |~S \wedge T & => & merge(S) \wedge merge(T) \\
         & |~\_ & => & T \\
  \end{array}
 \]

  \caption{Normalizing Types for \Fm.}
\label{fig:normalizationfn}

\end{figure}

\begin{lemma}
    For any type $T$, $nf(T)$ is in normal form.  Moreover, if $T$ is in normal form, $nf(T) = T$.
\end{lemma}

Moreover, types are equivalent to their normalized forms under the subtyping relationship.
\begin{lemma}
    $\Gamma~|~\Sigma \vdash nf(T) \sub T$ and $\Gamma~|~\Sigma \vdash T \sub nf(T)$.
\end{lemma}
\begin{proof}
    For one direction, note that $nf(nf(T)) = nf(T)$, and hence $nf(nf(T)) \sub nf(T)$.  Applying \ruleref{denormalize} allows us to show that $nf(T) \sub T$, as desired.  The other case follows
    by a symmetric argument.
\end{proof}

Not only does this allow us to simplify types to a normal form,
this  also allows us to state and prove canonical form lemmas and inversion lemmas, necessary 
for preservation and progress: Theorems \ref{theorem:preservation} and \ref{theorem:progress}.
Below we give examples for record types.  Similar lemmas exist and are mechanized
for function types and type-abstraction types as well.

\begin{lemma}[Inversion of Record Subtyping]
    If $S$ is a subtype of $\{f : T'\}$, and $S$ is in normal form,
    then at least one of its components is a type variable $X$ or a record type
    $\{f : S'\}$, where $\Gamma \vdash T' \sub S' \sub T'$.
\end{lemma}
\begin{lemma}[Canonical Forms for Records]
    If $v$ is a value and $\varnothing~|~\Sigma \vdash v : \{f : T\}$,
    then $v$ is a record and $f$ is a field of $v$ that maps to some location $l$.
\end{lemma}

\begin{lemma}[Inversion of Read-Only Record Subtyping]
    If $S$ is a subtype of $\readonly \{f : T'\}$, and $S$ is in normal form,
    then at least one of its components is a type variable $X$, read-only type variable $\readonly X$,
    a record type $\{f : S'\}$ where $\Gamma \vdash T' \sub S' \sub T'$, or a read-only record type 
    $\readonly \{f : S'\}$ where $\Gamma \vdash T' \sub S' \sub T'$.
\end{lemma}
\begin{lemma}[Canonical Forms for Read-Only Records]
    If $v$ is a value and $\varnothing~|~\Sigma \vdash v : \readonly \{f : T\}$,
    then $v$ is a record or a sealed record and $f$ is a field of $v$ that maps to some location $l$.
\end{lemma}

Note that normalization is necessary to state the inversion lemmas for read-only records, as $\readonly \{f : T'\}$,
$\readonly \readonly \{f : T'\}$, etc, give an infinite series of syntactically in-equivalent but semantically
equivalent types describing the same object -- a read-only record where field $f$ has type $T'$.

\subsubsection{Operational Safety}
\label{subsection:operational}
Operationally, we give small-step reduction semantics coupled with a store to \Fm in Figure \ref{fig:evaluation}.

\begin{figure}
  \judgement{Evaluation}{\fbox{$\langle s, \sigma\rangle \reduces \langle t, \sigma' \rangle$}}

  \begin{minipage}{0.45\textwidth}{\small
    \vspace{1.1em}
    \infax[\ruledefN{beta-v}{beta-v}]{
        \langle (\lambda x.t)(v), \sigma \rangle \reduces \langle t[x \mapsto v], \sigma \rangle
    }
    \vspace{1em}
    \infrule[\ruledefN{record-store}{record-store}]{
        l_i \notin \sigma
    }{
        \langle \{x_i : v_i\}, \sigma \rangle \reduces \langle \{x_i : l_i\}, (\sigma, l_1 : v_1, l_2 : v_2, \hdots) \rangle 
    }
    \vspace{1em}
    \infrule[\ruledefN{field}{field}]{
        l : v \in \sigma 
    }{
        \langle \{\hdots x : l \hdots\}.x, \sigma \rangle \reduces \langle v, \sigma \rangle
    }
    \vspace{1em}
    \vspace{1em}
    \infrule[\ruledefN{write-field}{write-field}]{
        l : v \in \sigma 
    }{
        \langle \{\hdots x : l \hdots\}.x = v', \sigma \rangle \reduces \langle v, \sigma[l \mapsto v'] \rangle
    }
    \vspace{1em}
    \infax[\ruledefN{beta-T}{beta-T}]{
        \langle (\Lambda (X\sub S).t)[T], \sigma \rangle \reduces \langle t[X \mapsto T], \sigma \rangle
    }}
  \end{minipage}\hfill
  \begin{minipage}{0.45\textwidth}{\small
    
    \infrule[\ruledefN{sealed-field}{sealed-field}]{
        l : v \in \sigma 
    }{
        \langle \highlight{(\seal \{\hdots x : l \hdots\})}.x, \sigma \rangle \reduces \langle \highlight{\seal v}, \sigma \rangle
    }
    \vspace{1em}
    \infax[\ruledefN{seal-elim-abs}{seal-elim-abs}]{
        \langle \highlight{\seal (\lambda x . t)}, \sigma \rangle \reduces \langle \lambda x . t, \sigma \rangle
    }
    \vspace{1em}
    \infax[\ruledefN{seal-elim-tabs}{seal-elim-tabs}]{
        \langle \highlight{\seal (\Lambda (X \sub S) . t)}, \sigma \rangle \reduces \langle \Lambda (X \sub S) . t, \sigma\rangle
    }
    \vspace{1em}
    \infax[\ruledefN{seal-elim-multiple}{seal-elim-multiple}]{
        \langle \highlight{\seal \seal v}, \sigma \rangle \reduces \langle \seal v, \sigma\rangle
    }
    \vspace{1em}
    \infrule[\ruledefN{context}{context}]{
        \langle s, \sigma \rangle \reduces \langle t, \sigma' \rangle
    }{
        \langle E[s], \sigma \rangle \reduces \langle E[t], \sigma' \rangle
    }}
    \end{minipage}

    \begin{center}
    \[
    \begin{array}{lcll}
        E & ::= & []~|~E(t)~|~v(E)~|~E[T] & \mbox{{\bf Evaluation Context}}\\
            & |   & \{x_0 : v_0, \hdots, x_i : E, x_{i+1} : t_{i+1}, \hdots\}\\
            & |   & E.x \\
            & |   & E.x = t~|~v.x = E \\
            & |   & \highlight{\seal E}
    \end{array}
    \]
    \end{center}

    \caption{Reduction rules for \Fm}
    \label{fig:evaluation}

    \end{figure}

Again, these rules are a straightforward extension of \Fsub with mutable boxes and records, with additional rules for
reducing sealed records.  To prove progress and preservation theorems, we 
additionally need to ensure that the store $\sigma$ itself is well typed in the context of some store typing
environment $\Sigma$ -- see rule \ruleref{store}.

The crux of preservation for \Fm is to show that {\tt seal}ed records are never given a non-read-only
type, so that the typing rule for reading from a mutable record -- \ruleref{record-elim} -- cannot 
be applied to {\tt seal}ed record values.
\begin{lemma}
    \label{lemma:typing-canonical-sealed}
    Suppose $\Gamma~|~\Sigma \vdash \seal r : T$ for some record $r$.  If $T$ is in normal form,
    then the components of $T$ are:
    \begin{itemize}
        \item The top type $\top$, or
        \item a read-only record type $\readonly \{f : T'\}$.
    \end{itemize}
\end{lemma}
From this key result we can show that preservation holds for \Fm.
\begin{theorem}[Preservation of \Fm]
    \label{theorem:preservation}
    Suppose $\langle s, \sigma \rangle \reduces \langle t, \sigma' \rangle$. 
    If $\Gamma~|~\Sigma \vdash \sigma$ and $\Gamma~|~\Sigma \vdash s : T$ for some type $T$,
    then there is some environment extension $\Sigma'$ of $\Sigma$ such that $\Gamma~|~\Sigma' \vdash \sigma'$ and
    $\Gamma~|~\Sigma' \vdash t : T$.
\end{theorem}

Conversely, values given a non-read-only record type must be an unsealed collection of references.
\begin{lemma}
    \label{lemma:canonical-record}
    Suppose $\varnothing~|~\Sigma \vdash v : \{f : T\}$ for runtime value $v$.  Then $v$ is an unsealed runtime record
    where field $f$ maps to some location $l$.
\end{lemma}

This lemma is needed to prove progress.

\begin{theorem}[Progress for \Fm]
    \label{theorem:progress}
    Suppose $\varnothing~|~\Sigma \vdash \sigma$ and $\varnothing, \Sigma \vdash s : T$.  Then either
    $s$ is a value or there is some $t$ and $\sigma'$ such that $\langle s, \sigma \rangle \reduces \langle t, \sigma' \rangle$.
\end{theorem}

\section{Static Immutability Safety}
\label{section:immutable}

Armed with Progress and Preservation, we can state immutability safety for full \Fm.
\Lm allows us to show that sealed records are never used to mutate their underlying
referenced values. \Fm shows that well-typed programs using seals do not get stuck. 
To prove immutability safety for \Fm, one problem still remains --
\Fm allows records that are not sealed to be given a read-only type.
We still need to show that records with such a type
are not used to mutate their values. In other words, we need to show
that records with a read-only type {\it could be} sealed,
and that the resulting program would execute in the same way.

We will do this by showing that,
given an original, well-typed \Fm program $s$, we can add {\tt seal}s to its read-only
subterms to obtain a new, well-typed \Fm program $t$, and furthermore that $t$ behaves the same way as $s$,
up to having additional {\tt seal}s in the resulting state.

The first step is to show that
{\tt seal}ing does not disturb the typing judgment
for terms.
\begin{lemma}
    \label{lemma:reseal}
    Suppose $\Gamma~|~\Sigma \vdash t : \readonly T$.  Then $\Gamma~|~\Sigma \vdash \seal t : \readonly T$.
\end{lemma}
\begin{proof}
    By \ruleref{seal}, $\Gamma~|~\Sigma \vdash \seal t : \readonly \readonly T$.  Then since $\readonly \readonly T <: \readonly T$,
    by \ruleref{sub}, $\Gamma~|~\Sigma \vdash \seal t : \readonly T$, as desired.
\end{proof}

From this, given a term $s$ and a typing derivation for $s$, $D = \Gamma~|~\Sigma \vdash s : T$, we can seal
those subterms of $s$ that are given a read-only type in $D$.
\begin{lemma}
    \label{lemma:seal-typing}
    Let $C$ be a term context with $n$ holes, and let $s=C[s_1, s_2, s_3, \hdots, s_n]$ be a term.
    Suppose $D$ is a typing derivation showing that $\Gamma~|~\Sigma \vdash s : T$.  Suppose also that
    $D$ gives each subterm $s_i$ of $s$ a type $\readonly T_i$.  Then $s' = s[\seal s_1, \seal s_2, \hdots, \seal s_n]$ has the following properties:
    \begin{enumerate}
        \item $s \leq s'$, and
        \item There exists a typing derivation $D'$ showing that $\Gamma~|~\Sigma \vdash s' : T$ as well.
    \end{enumerate}
\end{lemma}
\begin{proof}
    (1) is by definition.  As for (2), to construct $D'$,
    walk through the typing derivation $D$ showing that $\Gamma~|~\Sigma \vdash s : T$.
    When we reach the point in the typing derivation that shows that $s_i$
    is given the type $\readonly T_i$, note that $\seal s_i$ can also be given the type $\readonly T_i$
    by the derivation given by Lemma \ref{lemma:reseal}.  Replace the sub-derivation in $D$
    with the derivation given by Lemma \ref{lemma:reseal} to give a derivation in $D'$ for $\seal s_i$,
    as desired.
\end{proof}

This motivates the following definition.
\newcommand{\crest}{\texttt{crest}}
\begin{definition}
    Let $s$ be a term and let $D = \Gamma~|~\Sigma \vdash s : T$ be a typing derivation for $s$.
    Define $\crest(s,D)$ to be the term constructed from $s$ by replacing all subterms $s_i$ of $s$
    given a read-only type in $D$ by $\seal s_i$.
\end{definition}

A crested term essentially seals any sub-term of the original term
that is given a read-only type in a particular typing derivation.
By definition, for any term $s$ and typing derivation $D$ for $s$, we have $s \leq \crest(s,D)$.
Moreover, a crested term can be given the same type as its original term as well.

\begin{lemma}
    \label{lemma:crest-typing}
    Let $s$ be a term and let $D = \Gamma~|~\Sigma \vdash s : T$ be a typing derivation for $s$.
    Then $s \leq \crest(s,D)$, and there exists a typing derivation showing that $\Gamma~|~\Sigma \vdash \crest(s, D) : T$ as well.
\end{lemma}

Now by progress -- Theorem~\ref{theorem:progress} -- we have
that for any well typed term $s$ with typing derivation $D = \varnothing~|~\Sigma \vdash s : T$, its protected -- crested -- version $\crest(s,D)$ will also step.  By preservation -- Theorem~\ref{theorem:preservation} -- we have
that $\crest(s,D)$ either eventually steps to a value or runs forever, but never gets stuck.
It remains to relate the reduction steps of $\crest(s,D)$ to those of $s$, and specifically to show that if one reduces to some specific value and store, then the other also reduces to an equivalent pair of value and store.

We will do so by using the dynamic immutability safety properties proven in Section \ref{sec:dynamic}. 
\Fm satisfies the same sealing-equivalence properties as \Lm\ -- seals do not 
affect reduction, except perhaps by introducing other seals.  The following
are analogues of Lemmas
\ref{lemma:un-safety-value},
\ref{lemma:un-safety-step},
and
\ref{lemma:un-safety-step-exists}
for \Fm.

\begin{lemma}
    \label{lemma:safety-value}
    Let $v$ be a value, $\sigma_v$ be a store, $t$ be a term such that $v \leq t$, and $\sigma_t$ be
    a store such that $\sigma_v \leq \sigma_t$.
    
    If $\langle t, \sigma_t \rangle \reduces \langle t', \sigma_t'\rangle$
    then $v \leq t'$, $\sigma_v \leq \sigma_t'$, and $|t'| < |t|$. 
\end{lemma}
\begin{lemma}
    \label{lemma:safety-step}
    Let $s, t$ be terms such that $s \leq t$ and let $\sigma_s, \sigma_t$ be stores such that
    $\sigma_s \leq \sigma_t$.  If $\langle s, \sigma_s\rangle \reduces \langle s', \sigma_s'\rangle$
    and $\langle t, \sigma_t\rangle \reduces \langle t', \sigma_t'\rangle$ then:
    \begin{enumerate}
        \item Either $s \leq t'$, $\sigma_s \leq \sigma_t'$, and $|t'| < |t|$, or
        \item $s' \leq t'$ and $\sigma_s' \leq \sigma_t'$.
    \end{enumerate}
\end{lemma}
\begin{lemma}
    \label{lemma:safety-step-exists}
    Let $s, t$ be terms such that $s \leq t$ and let $\sigma_s, \sigma_t$ be stores such that
    $\sigma_s \leq \sigma_t$.  If $\langle t, \sigma_t\rangle \reduces \langle t', \sigma_t'\rangle$ then:
    \begin{enumerate}
        \item Either $s \leq t'$, $\sigma_s \leq \sigma_t'$, and $|t'| < |t|$, or
        \item There exists $s'$ and $\sigma_s'$ such that $\langle s, \sigma_s\rangle \reduces \langle s', \sigma_s'\rangle$, $s' \leq t'$ and $\sigma_s' \leq \sigma_t'$.
    \end{enumerate}
\end{lemma}

Stepping back, we can see using Lemma \ref{lemma:safety-step} that one step of $s$ to a term $s'$ corresponds to {\it finitely many} steps of $\crest(s,D)$; every step that $\crest(s,D)$ takes either removes a {\tt seal} or
corresponds to a reduction step that $s$ originally took. Hence $\crest(s,D)$ eventually steps to a term $t'$ such that $s' \leq t'$, preserving the desired equivalence of reduction between $s$ and $\crest(s,D)$.  The following is a generalization of the previous statement to two arbitrarily chosen well-typed terms $s$ and $t$ satisfying $s \leq t$.
\begin{lemma} Suppose $\varnothing, \Sigma \vdash \sigma_s$ and $\varnothing, \Sigma \vdash s : T$.
    Suppose $\langle s, \sigma_s \rangle \reduces \langle s', \sigma_s' \rangle$.
    For $\sigma_s \leq \sigma_t$, and $s \leq t$, such that $\Gamma, \Sigma \vdash \sigma_s$
    and $\Gamma, \Sigma \vdash t : T$, we have that $\langle t, \sigma_t \rangle \rreduces
    \langle t', \sigma_t' \rangle$ where $s' \leq t'$ and $\sigma_s' \leq \sigma_t'$.
    \label{lemma:progress-safety-step}
\end{lemma}
\begin{proof}

    From Theorem \ref{theorem:progress} we have that there exists a $t'$ and $\sigma_t'$
    that $\langle t, \sigma_t\rangle \reduces \langle t', \sigma_t'\rangle$.
    By Lemma \ref{lemma:safety-step} we have that either $s \leq t'$, $\sigma_s \leq \sigma_t'$, and
    $|t'| < |t|$,
    or that $s' \leq t'$ and $\sigma_s' \leq \sigma_t'$.  If $s' \leq t'$ and $\sigma_s' \leq \sigma_t'$
    we are done.  Otherwise, observe that since $|t'| < |t|$, a seal was removed.
    This can only occur a finite number
    of times, as $t$ and $t'$ have at most a finite number of seals, so we can simply loop until we obtain a $t'$ and $\sigma_t'$ such that $s' \leq t'$ and $\sigma_s' \leq \sigma_t'$. Note that Preservation -- Theorem \ref{theorem:preservation}
    allows us to do so as each intermediate step $t'$ can be given the same type $\Gamma~|~\Sigma \vdash t': T$.
\end{proof}

Finally, when $s$ eventually reduces to a value $v$, we can use Lemma \ref{lemma:safety-value} to show
that $\crest(s,D)$ reduces to a similar value $v'$ as well.  Again, the following is a generalization
of the previous statement to two arbitrarily chosen well-typed terms $s$ and $t$  satisfying $s \leq t$.
\begin{lemma}
    Suppose $\varnothing, \Sigma \vdash \sigma_s$ and $\varnothing, \Sigma \vdash s : T$
    such that $s$ eventually reduces to a value $v_s$ -- namely, that $\langle s, \sigma_e\rangle  \rreduces \langle v_s, \sigma_s'\rangle$ for some $\sigma_s'$.

    Then for any $t$ such that $s \leq t$ and $\varnothing, \Sigma \vdash t : T$,
    we have that $t$ eventually reduces to some value $v_t$,
    -- namely $\langle t, \sigma_e\rangle \rreduces \langle v_t, \sigma_t'\rangle$, such that
    $v_s \leq v_t$ and $\sigma_s' \leq \sigma_t'$. 
    \label{lemma:progress-safety}
\end{lemma}
\begin{proof}
    For each step in the multi-step reduction from $\langle s, \sigma_e\rangle \rreduces \langle v_s, \sigma_s'\rangle$ we can apply Lemma \ref{lemma:progress-safety-step} to show that
    $\langle t, \sigma_t\rangle$ eventually reduces to $\langle t', \sigma_t'\rangle$ where $v_s \leq t'$
    and $\sigma_s' \leq \sigma_t'$.  Now by Theorem \ref{theorem:progress} and Lemma ~\ref{lemma:safety-value} we have that either $t'$ is a value, in which case we are done, or that $\langle t', \sigma_t' \rangle$ steps to $\langle t'', \sigma_s' \rangle$ where $v_s \leq t''$.  Again, we can only
    take a finite number of steps of this fashion as the rule which reduces $t' \reduces t''$
    can only be one that removed a seal, so eventually we obtain a value $v_s$ such that $\langle t, \sigma_s \rangle \rreduces \langle v_t, \sigma_t'\rangle$ with $v_s \leq v_t$, and $\sigma_s' \leq \sigma_t'$, as desired.
    Again, note that Preservation -- Theorem \ref{theorem:preservation}
    allows us to do so as each intermediate step $t'$ can be given the same type $\Gamma~|~\Sigma \vdash t': T$.
\end{proof}

Now from Lemma~\ref{lemma:progress-safety} we obtain our desired immutability safety
results as a consequence -- namely, given a well-typed term $s$ 
that reduces to a value $v_s$, any references in $s$ with a $\readonly$ type are never actually
mutated, since they can be transparently sealed (which does not change the typing) to no ill effect.  Formally,
our main result is:
\begin{theorem}
    \label{lemma:progress-replace}
    Suppose $s$ is a term, $D = \varnothing~|~\Sigma \vdash s : T$ is a typing derivation for $s$,
    and let $\sigma_s$ be some initial store such that $\varnothing~|~\Sigma \vdash \sigma_s$.
    Then:
    \begin{itemize}
        \item $\crest(s, D)$ can be given the same type as $s$ -- $\varnothing~|~\Sigma \vdash crest(s,D) : T$.
    \end{itemize}

    Moreover, if $\langle s, \sigma_s\rangle \rreduces \langle v_s, \sigma_s'\rangle$, for some value $v_s$,
    then:
    \begin{itemize}
        \item $\crest(s, D)$ will reduce to a value $v_t$ -- $\langle crest(s, D), \sigma_e\rangle \rreduces \langle v_t, \sigma_t'\rangle$, such that
        \item $v_t$ and $\sigma_t'$ are equivalent to $v_s$ and $\sigma_s'$,
            modulo additional seals -- namely, that
            $v_s \leq v_t$ and $\sigma_s' \leq \sigma_t'$.
    \end{itemize}
\end{theorem}

Finally, it is useful to show that the converse result is also true; seals can be safely {\it removed} without affecting reduction.  First note that {\tt seal}s themselves can be transparently removed
without affecting the types assigned to the term.
\begin{lemma}
    \label{lemma:seal-remove-typing}
    Suppose $\Gamma~|~\Sigma \vdash \seal s : T$.  Then $\Gamma~|~\Sigma \vdash s : T$.
\end{lemma}

Moreover, the following analogue of Lemma \ref{lemma:un-erasure} holds in \Fm.
\begin{lemma}    
    \label{lemma:erasure}
    Suppose $s$ and $t$ are terms such that $s \leq t$.  If $\langle t, \sigma_t\rangle \rreduces \langle v_t, \sigma_t'\rangle$ for some value $v_t$,
    then for any $\sigma_s \leq \sigma_t$ we have $\langle s, \sigma_s\rangle \rreduces \langle v_s, \sigma_s'\rangle$ such that
    $v_s \leq v_t$ and $\sigma_s' \leq \sigma_t'$.
\end{lemma}

While Lemma~\ref{lemma:erasure} is enough to show when $s \leq t$, if $t$ reduces to a value then
so does $s$, we need Lemma~\ref{lemma:typed-erasure} to reason about the types of $s$ and $v_s$.
\begin{lemma}
    \label{lemma:typed-erasure}
        Suppose $s$ and $t$ are terms such that $s \leq t$.  If $\langle t, \sigma_t\rangle \rreduces \langle v_t, \sigma_t'\rangle$ for some value $v_t$,
    then for any $\sigma_s \leq \sigma_t$ we have $\langle s, \sigma_s\rangle \rreduces \langle v_s, \sigma_s'\rangle$ for some value $v_s$ such that
    $v_s' \leq v_s'$ and $\sigma_s' \leq \sigma_t'$.
    Moreover, $\Gamma~|~\Sigma \vdash s : T$ and $\Gamma~|~(\Sigma', \Sigma) \vdash v_s : T$ for some $\Sigma'$ as well.
\end{lemma}
\begin{proof}
    By Lemma \ref{lemma:seal-remove-typing} we can show that  $\Gamma~|~\Sigma \vdash s : T$.
    By Lemma \ref{lemma:erasure} we have that $v$ reduces to some value $v_s$.  By preservation --
    Theorem \ref{theorem:preservation} we have that $v_s$ has type $T$, as desired.
\end{proof}

\section{Mechanization}
Our mechanization of \Fm is based on the mechanization of \Fsub by \citet{10.1145/1328438.1328443}.
Our mechanization is a faithful model of \Fm as described in this paper except for one case.
To facilitate mechanization, reduction in our mechanization is done via explicit congruence rules in
each reduction rule instead of an implicit rule for reducing inside an evaluation context, similar to
how \citet{10.1145/1328438.1328443} originally mechanize \Fsub as well.

Proofs for all lemmas except for Theorem \ref{lemma:progress-replace} and Lemmas \ref{lemma:write-field-stuck},  \ref{lemma:seal-typing}, and 
\ref{lemma:crest-typing} have been mechanized using Coq 8.15 in the attached artifact.  Theorem \ref{lemma:progress-replace} and Lemmas
\ref{lemma:seal-typing}, \ref{lemma:crest-typing}, and  \ref{lemma:typed-erasure} have not been mechanized as they require computation
on typing derivations which is hard to encode in Coq as computation on {\tt Prop} cannot be reflected
into {\tt Set}.  Lemma \ref{lemma:write-field-stuck} has been omitted from our mechanization as it is hard to formally state, let alone prove, in a setting where reduction is done by congruence, though it almost follows intuitively from how the reduction rules are set up.

As the proofs of Lemmas \ref{lemma:safety-value}, \ref{lemma:safety-step},
\ref{lemma:safety-step-exists}, and \ref{lemma:erasure} do not rely on any extra structure present in \Fm over \Lm,
proofs for their \Lm analogues Lemmas \ref{lemma:un-safety-value}, \ref{lemma:un-safety-step},
\ref{lemma:un-safety-step-exists}, and \ref{lemma:un-erasure} have been omitted, as they can be recovered by erasing
the appropriate cases from their \Fm analogues.

\section{Related and Future Work}
\label{sec:related}
\subsection{Limitations -- Parametric Mutability Polymorphism}
Unlike other systems, \Fm does not support directly mutability polymorphism, neither through
a restricted {\tt @polyread} modifier as seen in \citet{10.1145/2384616.2384680}, nor through
explicit mutability variables as seen in \citet{10.1145/2384616.2384619}.   

This is a true limitation of \Fm, however, we note that it is possible to desugar parametric mutability polymorphism from a surface language into a core calculus like \Fm.
As \citet{10.1145/2384616.2384680} point out in their work, parametric mutability polymorphism can be desugared via {\it overloading},
noting that overloading itself can be dealt with in a surface language before desugaring into a base calculus, as seen before with Featherweight Java~\cite{10.1145/503502.503505}.

For example, consider the following top-level parametric function,
{\tt access}, which is parametric on mutability variable {\tt M}:
\begin{lstlisting}[language=Scala]
    def access[M](z: [M] Pair[Pair[Int]]): M Pair[Int] = { z.first }
\end{lstlisting}
This function can be rewritten instead as two functions with the same name {\tt access}, 
one taking in a regular, mutable pair, and one taking in a a readonly pair:
\begin{lstlisting}[language=Scala]
    def access(z: Pair[Pair[Int]]): Pair[Int] = { z.first }
    def access(@readonly z: Pair[Pair[Int]]): @readonly Pair[Int] = { z.first }
\end{lstlisting}

Nested and first-class functions are a little trickier but one can view a polymorphic, first-class
function value as a read-only record packaging up both overloads.
\begin{lstlisting}[language=Scala]
    {
        access: (z: Pair[Pair[Int]]) => { z.first },
        access: (@readonly z: Pair[Pair[Int]]) => { z.first }
    }
\end{lstlisting}

It would be interesting future work to see how one could add parametric mutability polymorphism to \Fm.

\subsection{Future Work -- Algorithmic Subtyping}
The subtyping rules of \Fm are fairly involved and it is difficult to see if an algorithmic subtyping system could
be devised.  We would conjecture that one could do so, using techniques from \citet{10.1145/3276482}'s integrated
subtyping work, but nonetheless algorithmic subtyping for \Fm remains an interesting and open problem.

\subsection{Viewpoint Adaptation}
Viewpoint adaptation has been used in reference immutability systems to denote the type-level adaptation
which is enforced to guarantee transitive immutability safety.
When a field $r.f$ is read from some record $r$,
the mutability of the resulting reference needs to be adapted from {\it both} the mutability
of $r$ and from the type of $f$ in the record itself.  While this notion of adaptation was known as
early as Javari \cite{10.1145/1103845.1094828}, the term ``viewpoint adaptation'' was first coined by \citet{10.1007/978-3-540-73589-2_3}.  They realized that this notion of adaptation could be
generalized to arbitrary qualifiers -- whether or not the type of a field read $r.f$ should
be qualified by some qualifier $@q$ should depend on whether or not $f$'s type is qualified and whether or
not $r$'s type is qualfied as well -- and used it to implement an {\it ownership} system for Java references
in order to tame {\it aliasing} in Java programs.

\subsection{Reference Immutability}
Reference immutability has long been studied in the context of existing object-oriented languages such 
as Java and C\#, and more recently has been studied in impure, functional languages like Scala.

\paragraph{{\bf roDOT} \cite{dort_et_al:LIPIcs:2020:13175}:}  roDOT extends the calculus of Dependent Object Types  \cite{amin2016essence} with support for reference immutability.  In their system, immutability
constraints are expressed through a type member field $x.M$ of each object, where $x$ is mutable
if and only if $M \leq \bot$, and $x$ is read-only if and only if $M \geq \top$.  Polymorphism in roDOT
is out of all reference immutability systems closest to how polymorphism is done in \Fm.  Type variables
quantify over full types, and type variables can be further restricted to be read-only as in \Fm.
Constructing a read-only version of a type, like how we use {\tt readonly} in \Fm, is done in roDOT by
taking an intersection with a bound on the type member $M$. For example, {\tt inplace\_map}
from before could be expressed in roDOT using an intersection type to modify immutability on the
type variable {\tt X}:
\begin{lstlisting}[language=Scala]
    def inplace_map[X](Pair[X]: pair, f: (X & {M :> Any}) => X): Unit
\end{lstlisting}
Dort et. al. also prove that roDOT respects immutability safety, but with different techniques than how
we show immutability safety in \Fm.  Instead of giving operational semantics with special forms
that guard references from being mutated, and relying on progress and preservation to imply static safety,
they take a different approach and show instead that values on the heap that change during reduction must be reachable by some statically-typed mutable reference in the initial program.  roDOT is a stronger system
than \Fm, as in particular mutabilities can be combined.  For example, one could write
a generic {\tt getF} function which reads a field {\tt f} out of any record that has {\tt f} as a field
polymorphic over {\it both} the mutabilities of the record {\tt x} and the field {\tt f}:
\begin{lstlisting}[language=Scala]
    def getF[T](x: {M: *, f : T}) : T & {M :> x.M} = x.f
\end{lstlisting}
Here, the return type of {\tt getF} will give the proper, tightest, viewpoint-adapted type for reading
{\tt x.f} depending on both the mutabilities of {\tt x} and {\tt f}.  This is not directly expressible
in \Fm and can only be expressed using overloading:
\begin{lstlisting}[language=Scala]
    def getF[T](x: @readonly {f : T}): @readonly T = x.f
    def getF[T](x: {f : T}) : T = x.f
\end{lstlisting}
However, in contrast, roDOT is significantly more complicated than \Fm.

\paragraph{{\bf Immutability for C\#} \cite{10.1145/2384616.2384619}:}  Of all the object calculi with
reference immutability the calculus of \citet{10.1145/2384616.2384619} is closest to that of roDOT in terms of flexibility.  Polymorphism is possible over {\it both} mutabilities and types in Gordon's system, but must be done separately; type variables instead quantify over base types that have
not been qualified with some immutability annotation, whether that be read-only or mutable.  The {\tt inplace\_map} function that we discussed earlier would be expressed with both a base-type variable as well as a mutability variable:
\begin{lstlisting}[language=Scala]
    def inplace_map[M, X](Pair[M X]: pair, f: @readonly X => M X): Unit
\end{lstlisting}
Like roDOT, Gordon's system also allows for mutability annotations to be combined in types, in effect
allowing viewpoint adaptation to be expressed at the type level using the mutability operator {\tt \textasciitilde>}.  
For example, {\tt getF} could be written as the following in Gordon's system:
\begin{lstlisting}[language=Scala]
    def getF[MS, MT, T, S <: {f : MT T}](x: MS S) : (MS ~> MT) T = x.f
\end{lstlisting}
Unlike roDOT however, which allows for inferences to be drawn about the mutability of the type {\tt 
    (T \& \{M :> x.M\}).M} depending on the bounds on $T$ and $S$, the only allowable judgment we
    can draw about {\tt MS \textasciitilde> MT} is that it can be widened to {\tt @readonly}.  We cannot conclude,
    for example, that {\tt MS \textasciitilde> MT <: M} in the following, even though both {\tt MS <: M}
    and {\tt MT <: M}:
\begin{lstlisting}[language=Scala]
    def getF[M, MS <: M, MT <: M, T, S <: {f : MT T}](x: MS S) : (MS ~> MT) T = x.f
\end{lstlisting}
Gordon et. al. also demonstrate the soundness and immutability safety of their system but
through an embedding into a program logic \cite{10.1145/2480359.2429104}. 

\paragraph{{\bf Javari} \cite{10.1145/1103845.1094828}:} Reference immutability was first modelled in the context
of Java; Javari is the earliest such extension.  In Javari's formalization,
Lightweight Javari, type variables {\tt X} stand in for either other type variables, class types, and {\tt readonly}-qualified class types.  Unlike roDOT and \Fm, in Lightweight Javari, type variables {\bf cannot} be further qualified by the {\tt readonly} type qualifier.  Lightweight Javari, however, does support parametric mutability polymorphism for class types, but does not support parametric mutability polymorphism directly on methods.  Instead, limited parametric mutability method polymorphism in Javari, denoted with the keyword {\tt romaybe}, is desugared using overloading into the two underlying methods handling the read-only case and the mutable case replacing {\tt romaybe} in the source.  Our earlier example, {\tt getF}, can be written using {\tt romaybe}
as follows:
\begin{lstlisting}[language=Java]
    class HasF<T> {
        T f;
        romaybe T getF() romaybe { return f; }
    }
\end{lstlisting}
However, this example is inexpressible in the core calculus Lightweight Javari, as {\tt @readonly T} is ill-formed.  As for safety, immutability safety is done in Lightweight Javari through a case analysis on how typed Lightweight Javari program terms can reduce. 
\cite{10.1145/1103845.1094828} claim that the soundness of Lightweight Javari reduces to showing the soundness of Lightweight Java, but no formal proof is given.

\paragraph{{\bf ReIm}: \cite{10.1145/2384616.2384680}:} ReIm simplifies Javari to enable fast, scalable mutability
inference and analysis.  Like Javari, ReIm supports two type qualifiers -- {\tt readonly} and {\tt polyread},
where {\tt readonly} marks a read-only type and {\tt polyread} is an analogue of {\tt romaybe} from Javari.
Like Lightweight Javari, and unlike roDOT and \Fm, ReIm restricts how qualifiers interact with generics.  ReIm's polymorphism model is similar to that of \citet{10.1145/2384616.2384619} -- type variables range over unqualified types.  However, ReIm has no mechanism for mutability polymorphism, and therefore {\tt getF} cannot be written in ReIm at all.  Unlike other related work, neither soundness nor immutability safety is proven to hold for ReIm.

\paragraph{{\bf Immutability Generic Java:} \cite{10.1145/1287624.1287637}:} Immutability Generic Java is a scheme for expressing immutability using Java's existing generics system.  The type {\tt List<Mutable>} denotes
a {\it mutable} reference to a List, whereas the type {\tt List<Readonly>} denotes a {\it read-only} reference to a list.  Viewpoint adaptation is not supported, and transitive immutability must be explicitly opted into.  For example, in the following snippet, the field {\tt value} of {\tt C} is {\it always} mutable.  Transitive
immutability must be explicitly opted into by instantiating {\tt List} with the immutability parameter {\tt ImmutOfC}.
\begin{lstlisting}[language=Java]
    class C<ImmutOfC> {
        List<Mutable /* ImmutOfC for transitivity */, Int> value;
    }
\end{lstlisting}
Moreover, transitive immutability cannot be expressed at all over fields given a generic type.
Type variables by the nature of how immutability is expressed in IGJ range over fully qualified types,
and there is no mechanism for re-qualifying a type variable with a new immutability qualifier.  For example,
the mutability of {\tt value} in any {\tt Box} below depends {\it solely} on whether or not {\tt T} is mutable.
Hence the {\it value} field of a {\tt Box} is mutable even if it was read through a read-only {\tt Box} reference -- that is, a reference of type {\tt Box<ReadOnly>}.
\begin{lstlisting}[language=Java]
    class Box<ImmutOfBox, T> {
        T value;
    }

    Box<Readonly, List<Mutable,Int>> b = new Box(...)
    b.value.add(10); // OK -- even though it mutates the underlying List.
\end{lstlisting}

\subsection{Languages with Immutability Systems}
Finally, some languages have been explicitly designed with immutability in mind.

\paragraph{{\bf C++}:} {\tt const}-qualified methods and values provide limited viewpoint adaptation.
Reading a field from a {\tt const}-qualified object returns a {\tt const}-qualified field, and C++
supports function and method dispatching based on the {\tt const}ness of its arguments~\cite{DBLP:books/daglib/0019344}.  Mutability
polymorphism is not explicitly supported but can be done with a combination of templates and overloading.
\begin{lstlisting}[language=C++]
    struct BoxedInt {
        int v{0};  
    };
    template<typename T> struct HasF<T> {
        T f;
        T& getF() { return f; }
        const T& getF() const { return f; }
    }
    
    const HasF<BoxedInt> x;
    x.getF() // Calls const qualified getF()
    const BoxedInt& OK = x.f; // OK, as x.f is of type const BoxedInt.
    BoxedInt& Bad = x.f; // Bad, discards const-qualifier.
\end{lstlisting}
In this example a C++ compiler would disallow {\tt Bad} because the type of {\tt x.f} has been adapted
to a l-value of {\tt const BoxedInt}.  However, viewpoint adaptation does not lift to reference
or pointer types in C++.  For example, if instead we had a pointer-to-{\tt T} in {\tt HasF}:
\begin{lstlisting}[language=C++]
    template<typename T> struct HasF<T> {
        T* f;
    }
    
    BoxedInt b{5};
    const HasF<BoxedInt> x{&b};
    BoxedInt* NotGreat = x.f; // OK, as x stores a constant pointer to a mutable BoxedInt
    NotGreat->v = 10; // Modifies b!
\end{lstlisting}
C++'s limited viewpoint adaptation gives {\tt x.f} the type {\tt BoxedInt * const},
which is a constant pointer to a mutable {\tt BoxedInt}, not the type {\tt BoxedInt const * const},
which would be a constant pointer to a {\it constant} {\tt BoxedInt}.  This allows
the underlying field to be mutated.

\paragraph{{\bf D}:} In contrast to C++, where {\tt const} becomes useless for pointer and reference
fields, D supports full reference immutability and viewpoint adaptation with a {\it transitive} {\tt const} extended to work for pointer and reference types~\cite{10.1145/3386323}.  
Again, mutability polymorphism is not directly supported but can be encoded with D's compile-time meta-programming system.

\paragraph{{\bf Rust}:} In Rust, references are either {\it mutable} or {\it read-only}, and only
one {\it mutable} reference can exist for any given value.  Read-only references are transitive,
like they are in \Fm, roDOT, and other reference immutability systems, and unlike C++.  Here,
in this example, we cannot write to {\tt s3.f} as it {\tt s3} is an read-only reference to {\tt s2},
even though {\tt s2.f} has type {\tt \&mut String}.
\begin{lstlisting}[language=Rust]
    struct HasF<T> {
        f: T
    }
    
    fn main() {
        let mut s1 = String::from("hello");
        let s2 = HasF { f: &mut s1 };
        s2.f.push_str("OK");
        let s3 = &s2;
        s3.f.push_str("BAD");
    }
\end{lstlisting}

Unlike other languages, though, the mutability of a reference is an intrinsic property of the reference type itself.
Instead of having a type operator {\tt readonly} that, given a reference type {\tt T}, creates a read-only
version of that reference type, Rust instead defines {\tt \&} and {\tt \&mut}, type operators
that, given a type {\tt T}, produce the type of a read-only reference to a {\tt T} and
the type of a mutable reference to a {\tt T}, respectively.  Here, in the following example,
{\tt s1} is a {\tt String}, {\tt s2} is a mutable reference to a {\tt s1} -- {\tt \&mut String}, and {\tt s3} is a read-only reference to {\tt s2} -- {\tt \& (\&mut String)}, where all three of {\tt s1}, {\tt s2}, and {\tt s3} are stored at distinct locations in memory.

\begin{lstlisting}[language=Rust]
    let s1 = String::from("hello");
    let mut s2 = &s1;
    let s3 = &s2;
\end{lstlisting}

As such, in Rust, one cannot create a read-only version of an existing reference type.
This makes higher-order functions over references that are polymorphic over mutability, 
like {\tt inplace\_map} from above, inexpressible in Rust.  However, if we instead had a {\tt Pair}
that owned its elements, we could write the following version of {\tt inplace\_map}:
\begin{lstlisting}[language=Rust]
    struct Pair<T> {
        fst: T,
        snd: T
    }
    
    fn inplace_map<T>(p: &mut Pair<T>, f: fn (&T) -> T) {
        p.fst = f(&p.fst);
        p.snd = f(&p.snd);
    }
\end{lstlisting}
Note, though, that in this setting, the elements {\tt p.fst} and {\tt p.snd} are embedded in the pair {\tt p}
and owned by it.

\subsection{Type Qualifiers and Polymorphism}
\citet{10.1145/301618.301665} formalize
a system for enriching types with qualifiers with support for polymorphism over both ground, unqualified
types and qualifiers themselves.  In this setting, {\tt readonly} can be viewed as a type qualifier,
similar to how C++'s {\tt const} can be viewed as a qualifier in \cite{10.1145/301618.301665}.  The
resulting calculus which arises is similar to the calculus of \cite{10.1145/2384616.2384619} restricted
only to reference immutability qualifiers.

\subsection{Contracts}
Our approach to sealing references is similar to and was inspired by practical programming experience 
with Racket contracts -- \cite{10.1145/2384616.2384685}.  
{\tt Seal}ing, in particular, can be viewed as attaching a {\it chaperone} contract which
raises an exception whenever the underlying chaperoned value is written to, and attaches
fa similar chaperone to every value read out of the value.  For example, a
dynamic reference immutability scheme for Racket vectors could be implemented with the following
chaperone contract:
\begin{lstlisting}[language=Scheme]
    (define (chaperone-read vec idx v)
      (seal v))
    (define (chaperone-write vec idx v)
      (error 'seal "Tried to write through an immutable reference."))
    
    (define (seal v)
      (cond
        [(vector? v) (chaperone-vector vec chaperone-read chaperone-write))
        [else v]))
\end{lstlisting}
Strickland et. al. prove that chaperones can be safely erased without changing the behaviour
of the underlying program when it reduces to a value.  Our results on dynamic safety, Lemmas~\ref{lemma:un-safety-value},~\ref{lemma:un-safety-step}, and~\ref{lemma:un-safety-step-exists} can be viewed as an analogue of \cite[Theorem 1]{10.1145/2384616.2384685} specialized to reference immutability.  In this setting, our static
immutability safety results show that a well-typed program will never raise an error by writing to a chaperoned
vector.

\section{Conclusion}
We contributed a {\it simple} and {\it sound} treatment of reference immutability in \Fsub.
We show how a simple idea, {\it sealing} references, can provide dynamic immutability safety guarantees
in an untyped context -- \Lm -- and how soundness and \Fsub-style polymorphism can be recovered in
a typed calculus \Fm which builds on both \Lm and \Fsub.
Our hope is to enable reference immutability systems in functional languages via this work, by giving simple soundness foundations in a calculus (\Fsub) which underpins many impure functional languages today.

\begin{acks}
We thank Yaoyu Zhao for his interesting discussions on reference immutability.
We thank Alexis Hunt and Hermann (Jianlin) Li for their useful feedback on early drafts of this work.
This work was partially supported by the Natural Sciences and Engineering Research Council of Canada and
by an Ontario Graduate Scholarship.  No seals were clubbed in the creation of this paper.
\end{acks}

\label{sec:conclusion}

\bibliography{main}
\end{document}